\input ./style/arxiv-general.cfg
\documentclass[aoas,MSNbibl,nameyear,dvips]{arximspdf}
\makeatletter
   \@ifpackageloaded{graphicx}{}{\usepackage{graphicx}}
\makeatother

\doi{10.1214/15-AOAS847}
\volume{9}
\issue{3}
\pubyear{2015}
\firstpage{1278}
\lastpage{1297}
\docsubty{FLA}

\makeatletter
\def\argmin{\mathop{\operatorname{argmin}}}
\newtheorem{theorem}{Theorem}
\makeatother

\begin{document}
\begin{frontmatter}

\title{A new framework for Euclidean summary statistics in the neural
spike train space}
\runtitle{Euclidean summary statistics in the neural spike train space}

\begin{aug}
\author[A]{\fnms{Sergiusz}~\snm{Wesolowski}\corref{}\ead[label=e1]{wesserg@gmail.com}},
\author[B]{\fnms{Robert J.}~\snm{Contreras}\ead[label=e2]{contreras@psy.fsu.edu}}
\and
\author[C]{\fnms{Wei}~\snm{Wu}\ead[label=e3]{wwu@stat.fsu.edu}}
\runauthor{S. Wesolowski, R. J. Contreras and W. Wu}
\affiliation{Florida State University}
\address[A]{S. Wesolowski\\
Department of Mathematics\\
Florida State University\\
Tallahassee, Florida 32306-4510\\
USA\\
\printead{e1}}
\address[B]{R. J. Contreras\\
Department of Psychology\\
Florida State University\\
Tallahassee, Florida 32306-4301\\
USA\\
\printead{e2}}
\address[C]{W. Wu\\
Department of Statistics\\
Florida State University\\
Tallahassee, Florida 32306-4330\\
USA\\
\printead{e3}}
\end{aug}

%
\received{\smonth{6} \syear{2014}}
%
\revised{\smonth{5} \syear{2015}}

%
\begin{abstract}
Statistical analysis and inference on spike trains is one of the
central topics in the neural coding.
It is of great interest to understand the underlying structure of given
neural data. Based on
the metric distances between spike trains, recent investigations have
introduced the notion
of an \textit{average} or \textit{prototype} spike train to
characterize the template pattern in
the neural activity. However, as those metrics lack certain Euclidean
properties, the
defined \textit{averages} are nonunique, and do not share the
conventional properties of
a mean. In this article, we propose a new framework to define the mean
spike train
where we adopt a Euclidean-like metric from an $L^p$ family. We
demonstrate that
this new mean spike train properly represents the average pattern in
the conventional
fashion, and can be effectively computed using a theoretically-proven convergent
procedure. We compare this mean with other spike train averages and
demonstrate its
superiority. Furthermore, we apply the new framework in a recording
from rodent
geniculate ganglion, where background firing activity is a common issue
for neural coding.
We show that the proposed mean spike train can be utilized to remove
the background noise and improve decoding performance.
\end{abstract}

%
\end{frontmatter}

\section{Introduction}\label{sec1}
Neural spike trains are often called the language of the brain and are
the focus of many investigations in computational neuroscience. Due to
the stochastic nature of the spike discharge record, probabilistic and
statistical methods have been
extensively investigated to examine the underlying firing patterns
[\citet{Rieke97,Brown02,Kass05,Box78,Kass01}]. However, these methods
focus only on parametric representations at each
given time and therefore can prove to be limited in data-driven
problems in the space of spike
trains.

Alternative approaches for analyzing spike train data are based on
metricizing the spike train space.
Over the past two decades, various methods have been
developed to measure distances or dissimilarities between spike
trains, for example, the distances in discrete state space, discrete time
models [\citet{Lim94,Macleod98,Rieke97}], those in discrete state space,
continuous time models [\citet{Victor96,Aronov03a,Aronov04,Victor07,WuSrivastavaJCNS11}], those in continuous state space,
continuous time models [\citet{VanRossum01,Houghton08,Houghton09}],
and a number of others [\citet{Schreiber03,Kreuz07,QuianQuiroga02,Hunter03,Paiva09b}].

An ongoing pursuit of great interest in computational neuroscience is
defining an average that can represent tendency of a set of spike
trains. What follows is the problem of defining basic summary
statistics reflecting the intuitive properties of the mean and the
variance, which are crucial for further statistical inference methods.
In particular, to make the first-order statistic, mean, convenient for
constructing new framework and inference methods, it should satisfy the
following properties:
\begin{longlist}[3.]
\item[1.] The mean is uniquely defined in a certain framework.
\item[2.] The mean is still a spike train.
\item[3.] The mean represents the conventional intuition of average like in
the Euclidean space.
\item[4.] The mean depends on exact spike times only, and is independent of
the recording time period.
\item[5.] The mean can be computed efficiently.
\end{longlist}
Property 3 can be described as follows: given a set of $N$ spike trains
with each having $K$ spikes, we denote these trains using vectors $\{
(x_{i1}, \ldots, x_{iK})\}_{i=1}^N$. Then the mean spike train is
expected to resemble $\frac{1}{N}\sum_{i=1}^N (x_{i1}, \ldots, x_{iK})$.

In \citet{vpk97}, the authors considered a ``consensus'' spike train,
which is the centroid of a spike train set (under the Victor--Purpura
metric). This idea was further generalized in \citet{Diez2012} to a
``prototype'' spike train which does not have to belong to the given
set of spike trains, but its spike times are chosen from the set of all
recorded spike times. Recently, a notion of an ``average'' based on
kernel smoothing methods was introduced in \citet{JMN13}. In
\citeauthor{WuSrivastavaJCNS11} (\citeyear{WuSrivastavaJCNS11,WuSrivastavaJCNS13}), the authors proposed an
elastic metric on inter-spike intervals to define a mean directly in
the spike train space. However, none of these approaches satisfies the
5 desirable properties listed above, and therefore may result in
limited use in practical applications.

In this article, we propose a new framework for defining the mean spike
train. We adopt a recently developed metric related to an $L^p$ family
with $p \ge1$, which inherits desirable properties in the special case
of $p=2$ [\citet{Dubbs10}]. This metric is a direct generalization of
the Victor--Purpura metric, and we refer to it as a GVP (Generalized
Victor--Purpura) metric. We will demonstrate that this new mean spike
train satisfies all aforementioned 5 properties. In particular, the new
framework is the only one (over all available methods) that has
desirable Euclidean properties on the given spike times.
Such properties are crucial for the definition of summary statistics
such as the mean, variance, and covariance in the spike train space.
In general, these 5 properties assure intuitiveness of the summary
statistics, as well as efficiency in their estimation.
In contrast, previous methods have issues such as nonuniqueness,
dependence on model assumptions, or more complicated computations, and
therefore do not result in the same level of performance (see the
detailed comparison in Section~\ref{subsec:comparison}).

One direct application of the mean spike train is in neural decoding in
the rodent peripheral gustatory system [\citet{WuPLOSONE13}]. The
neural data was recorded from single cells in geniculate ganglion, as
the spiking activity in these neurons modulated with respect to
different taste stimuli on the tongue. It is commonly known that
spontaneous spiking activity can be observed even if only artificial
saliva is applied. Thus, the neural response is actually a mixture of a
background activity and a taste-stimulus activity. In this article we
demonstrate using simulation as well as real data that the proposed new
framework can be used to remove the background activity, which leads to
improvement in decoding performance.

In Section~\ref{sec:methods} we define the new framework by
introducing the mean spike train under the GVP metric, and provide an
efficient algorithm to estimate it. In Section~\ref{sec:application}
we extend this framework by developing a statistical approach for noise
removal and apply the method to the experimental data. We then discuss
the merits of the new framework in Section~\ref{sec:discussion}.
Finally, in the \hyperref[appen]{Appendix}, we provide mathematical details on the
convergence of the mean estimation algorithm.

\section{New framework}
\label{sec:methods}

Before we turn to describing the methods, it is necessary to set up a
formal notation in the spike train space.
\subsection{Notation}
\label{subsec:notation}

Assume $S$ is a spike train with spike times $0 < s_1 < s_2 < \cdots<
s_M < T$, where $[0, T]$ denotes a recording time domain. We denote
this spike train~as
\[
S=(s_j)_{j=1}^{M}=(s_1,
s_2, \ldots, s_{M}).
\]
We define the
space of all spike trains containing $M$ spikes to be ${\mathcal S}_M $
and the space of all spike trains to be ${\mathcal S} =
\bigcup_{M=0}^{\infty} {\mathcal S}_M$. This can be equivalently
described as a space of all bounded, finite, increasing sequences.

A time warping on the spike times (or inter-spike intervals) has been
commonly used to measure distance between two spike trains [\citet
{Victor96,Dubbs10,WuSrivastavaJCNS11}]. Let $\Gamma$ be the set of
all time-warping functions,
where a time warping is defined as an orientation-preserving
diffeomorphism of the domain $[0, T]$. That; that is,
\[
\Gamma= \bigl\{\gamma: [0, T] \rightarrow[0, T] | \gamma(0) = 0, \gamma(T) = T,
0 < \dot \gamma(t) < \infty \bigr\}.
\]
It is easy to verify that $\Gamma$ is a \textit{group} with the operation
being the composition of functions. By applying
$\gamma\in\Gamma$ on a spike train $S=(s_j)_{j=1}^{M}$, one obtains
a warped spike train
$\gamma(S) = (\gamma(s_j))_{j=1}^{M}$.

\subsection{GVP metric}
In \citet{Dubbs10}, a new spike train metric was introduced with
parameter $p \in[1, \infty)$. This metric is a direct generalization
of the classical Victor--Purpura (VP) metric (VP is a special case when
$p=1$), and we refer to it as the Generalized
Victor--Purpura (GVP) metric. In particular, when $p=2$, this metric
resembles a Euclidean $L^2$ distance.

Assume that $X = (x_i)_{i=1}^M$ and $Y = (y_j)_{j=1}^N$ are two\vspace*{2pt} spike
trains in
$[0, T]$.
For $ \lambda(> 0)$, the GVP metric between $X$ and $Y$ is given in
the following form:
\begin{equation}
d_{\mathrm{GVP}}[\lambda] (X,Y) = \min_{\gamma\in\Gamma}
\biggl(E_{\mathrm{OR}} \bigl(X, \gamma(Y) \bigr) + \lambda^2 \sum
_{\{i,j: x_i = \gamma(y_j)\}} (x_i - y_j)^2
\biggr)^{1/2} , \label{eq:gvp}
\end{equation}
where $E_{\mathrm{OR}}(\cdot, \cdot)$ denotes the cardinality of the Exclusive
OR (i.e., union minus intersection) of two sets. That is, $E_{\mathrm{OR}}(X,
\gamma(Y))$ measures the number of unmatched spike times between $X$
and $\gamma(Y)$ and can be computed as
\[
E_{\mathrm{OR}} \bigl(X, \gamma(Y) \bigr) = M + N - 2\sum
_{i=1}^M \sum_{j=1}^N
{\mathbf1}_{\{
\gamma(y_j) = x_i\}},
\]
where ${\mathbf1}_{\{\cdot\}}$ is an indicator function. The constant
$\lambda( > 0)$ is the
penalty coefficient. We emphasize that $d_{\mathrm{GVP}}$ is a proper metric,
that is, it satisfies \textit{positive definiteness}, \textit{symmetry}, and
\textit{the triangle inequality}. It shares a lot of similarities with the
classical $L^2$ norm.

Similarly to the result in \citet{WuSrivastavaJCNS13}, one can show
that the optimal time warping between two spike trains $X =
(x_i)_{i=1}^M$ and $Y = (y_j)_{j=1}^N$ must be a strictly increasing,
piece-wise linear function, with nodes mapping from points in $Y$ to
points in $X$. Based on this fact, a dynamic programming algorithm was
developed to compute the distance $d_{\mathrm{GVP}}$ with the computational cost
of the order of $O(MN)$. Using the bipartite graph matching theory,
another efficient algorithm was also developed to compute $d_{\mathrm{GVP}}$ in
the cost of $O(MN)$ [\citet{Dubbs10}].


\subsection{Definition of the summary statistics and their properties}
\label{subsec:definition}

Conventional statistical inferences in the Euclidean space are based on
basic quantities such as mean and variance. For statistical inferences
in the spike train space, we can analogously use a Euclidean spike
train metric to define these summary statistics as follows.

For a set of spike trains $S_1, S_2, \ldots,
S_K \in\mathcal{S}$ where the corresponding numbers of spikes
are $n_1, n_2, \ldots, n_K$ (arbitrary nonnegative integers),
respectively, we define their sample
\textit{mean} using the classical Karcher mean [\citet{Karcher77}] as follows:
\begin{equation}
S^* = \argmin_{S \in\mathcal{S}} \sum_{k=1}^K
d_{\mathrm{GVP}}[\lambda ](S_k, S)^2. \label{eq:mean}
\end{equation}
When the mean spike train $S^*$ is known, the associated
(scalar) sample variance, $\sigma^2$, can be defined in the following form:
\begin{equation}
\sigma^2 = \frac{1}{K-1}\sum_{k=1}^K
d_{\mathrm{GVP}}[\lambda] \bigl(S_k, S^* \bigr)^2.
\label{eq:var}
\end{equation}
The computation of this variance is straightforward, and the main
challenge is
in computing the mean spike train for any $\lambda (>0)$.

Before we move on to the computational methods for the summary
statistics, we list several basic theoretical properties of the mean
spike trains using the $d_{\mathrm{GVP}}$ metric. The proofs are omitted here to
save space:
\begin{longlist}[3.]
\item[1.] The optimal time warping between two spike trains must be a
continuous, increasing, and piece-wise linear function between subsets
of spike times in these two trains.

\item[2.] Let spike trains $X = (x_i)_{i=1}^M, Y = (y_i)_{i=1}^M \in
\mathcal{S}_M$ be defined on $[0, T]$. If $\lambda^2 <
1/(MT^2)$, then
\[
d_{\mathrm{GVP}}[\lambda](X,Y) = \Biggl(\lambda^2 \sum
_{i=1}^{M}(x_i - y_i)^2
\Biggr)^{1/2} .\label{eq:lm2dis}
\]
\item[3.] Assume a set of spike trains $S_1,
S_2, \ldots, S_K \in\mathcal S$ with $n_1, n_2, \ldots, n_K$ spikes,
respectively, and let $N_{\max}= \max(n_1, n_2, \ldots, n_K)$. If
$\lambda^2 < 1/\break (KN_{\max}T^2)$, then the number of spikes in the mean
train is the \textit{median} of $\{n_k\}_{k=1}^K$.

\item[4.] Let spike trains $S_1, \ldots,
S_K \in\mathcal{S}_M$. If $\lambda^2 < 1/(KMT^2)$, then the mean
spike train has a conventional closed-form:
\[
\frac{1}{K}\sum_{k=1}^{K}
S_k.
\]
\end{longlist}

\subsection{Computation of the mean spike train}
\label{sec:comp_mean}
To compute the mean spike train $S^*$ under the GVP metric, we need to
estimate two unknowns: (1) the number of
spikes $n$, and (2) the placements of these spikes in $[0,T]$. For a
general value of $\lambda> 0$, neither the matching term nor the
penalty term is dominant,
and therefore we cannot identify the number of spikes in the mean
before estimating
their placements [\citet{WuSrivastavaJCNS11}]. A key challenge is that
we need to update the number of spikes in the algorithm. In this
article, we propose a general algorithm to estimate the mean spike train.
We initialize the number of spikes in the mean spike train equal to the maximum
of $\{n_1, n_2, \ldots, n_K\}$, and then adjust this number during the
iterations.
We present, here, how the Karcher mean in equation (\ref{eq:mean}) can
be efficiently computed using a convergent procedure.

\subsubsection{Algorithm}
Assume that we have a set of $K$ spike\vspace*{1pt} trains, $S_1, \ldots, S_K$ with
$n_1, n_2, \ldots, n_K$ spikes, respectively. Denote $S_k =
(s_i^k)_{i=1}^{n_k}$ and $S = (s_i)_{i=1}^{n}$. Then the sum of
squared distances in equation (\ref{eq:mean}) is
\begin{eqnarray}\label{eq:sumdubbs}
&& \sum_{k=1}^K d_{\mathrm{GVP}}[\lambda]
\bigl(S^k, S \bigr)^2
\nonumber\\[-8pt]\\[-8pt]\nonumber
&&\qquad  = \sum_{k=1}^K
\min_{\gamma
\in\Gamma} \biggl( E_{\mathrm{OR}} \bigl[S^k,
\gamma(S) \bigr] + \lambda^2 \sum_{\{i,j:
s_i^k = \gamma(s_j)\}}
\bigl(s^k_i - s_j \bigr)^2
\biggr).
\end{eqnarray}
We develop here an iterative procedure to minimize $\sum_{k=1}^K
d_{\mathrm{GVP}}[\lambda](S^k, S)^2$ (as a~function of $S$) and estimate the
optimal $S^*$. This new algorithm has four main steps in each
iteration: Matching, Adjusting, Pruning, and Checking, and we refer to
it as the MAPC algorithm. In particular, the Adjusting step corresponds
to the Centering step in the MCP-algorithm in \citet
{WuSrivastavaJCNS13}; in contrast to the nonlinear warping-based
Centering-step, the Adjusting step utilizes the Euclidean property and
updates the mean spike train in an efficient linear fashion. The
Checking step is mainly used to avoid underestimating the number of
spikes in the mean. This step adds one spike into the current mean and
checks if such an addition results in a better mean (i.e., smaller mean
squared distances). In contrast, such a problem is not addressed in the
MCP algorithm.\vspace*{6pt}

\textit{Matching--Adjusting--Pruning--Checking (MAPC) Algorithm}:
\begin{enumerate}
\item[1.] Let $n = \max\{n_1, n_2, \ldots, n_K\}$. (Randomly) set initial
times for the $n$
spikes in $[0, T]$ to form an initial $S$.
\item[2.] \textit{Matching step}: Use the dynamic programming procedure
[\citet{WuSrivastavaJCNS11}] to find the optimal matching $\gamma^k$
from $S$ to $S^k, k = 1, \ldots, K$. That~is,
\begin{equation}
\gamma^k = \argmin_{\gamma\in\Gamma} \biggl( E_{\mathrm{OR}}
\bigl[S^k, \gamma (S) \bigr] + \lambda^2 \sum
_{\{i,j: s_i^k = \gamma(s_j)\}} \bigl(s^k_i - s_j
\bigr)^2 \biggr).
\end{equation}
\item[3.] \textit{Adjusting step}:
\begin{longlist}[(a)]
\item[(a)] For $k = 1, \ldots, K, j = 1, \ldots, n$, define
\[
r_j^k = \cases{ s_i^k, &\quad
if $\exists i \in\{1, \ldots, n_k\}$, s.t. $\gamma^k(s_j)
= s_i^k$,
\cr
s_j, &\quad otherwise.}
\]
\item[(b)] Denote $R_k = (r_1^k, \ldots, r_n^k), k = 1, \ldots, K$.
Then we update the mean spike train to be
$ \tilde S = \frac{1}{K}\sum_{i=1}^K R_i$.
\end{longlist}

\item[4.] \textit{Pruning step}: Remove spikes from the proposed mean
$\bar S$ that are matched less than $K/2$ times.
\begin{longlist}[(a)]
\item[(a)] For each $j = 1, \ldots, n$, count the number of times
$s_j$ appears in $\{\gamma^k(S^k)\}_{k=1}^K$. That is,
$ h_j = \sum_{k=1}^N 1_{s_j \in{\gamma^k(S^k)}} $.
\item[(b)] Remove $s_j$ from $\tilde S$ if $h_j \le K/2, j = 1, \ldots
, n$, and denote the updated mean spike train as $\tilde S^*$. Then
\[
\tilde S^* = \{ s_j \in\tilde S | h_j > K/2\}.
\]
\end{longlist}
\item[5.]  \textit{Checking step}: To avoid being stuck in a local minimum,
we check if an insertion or/and deletion of a specific spike can
improve the mean estimation:
\begin{longlist}[(a)]
\item[(a)] Let $\hat{S}^*$ be $\tilde S^*$ except one spike with the
minimal number of appearances (randomly chosen if there are multiple
spikes at the minimum) in the Pruning step. Then, update the mean as
\[
S^{**} = \cases{ \hat S^*, &\quad if $\displaystyle\sum
_{k=1}^K d_{\mathrm{GVP}}[\lambda]
\bigl(S^k, \hat S^* \bigr)^2 < \sum
_{k=1}^K d_{\mathrm{GVP}}[\lambda]
\bigl(S^k, \tilde S^* \bigr)^2$,
\cr
\tilde S^*, &\quad
otherwise.}
\]
\item[(b)] Let $\hat S^{**}$ be the current mean $S^{**}$ with one
spike inserted at random within $[0, T]$. Then update the mean as
\[
S^{***} = \cases{ \hat S^{**}, &\quad if $\displaystyle\sum
_{k=1}^K d_{\mathrm{GVP}}[\lambda]
\bigl(S^k, \hat S^{**} \bigr)^2 < \sum
_{k=1}^K d_{\mathrm{GVP}}[\lambda]
\bigl(S^k, S^{**} \bigr)^2$,
\cr
\tilde
S^{**}, &\quad otherwise.}
\]
\end{longlist}

\item[6.] \textit{Mean update}: Let $S = S^{***}$ and $n$ be the number of spikes in $S$.

\item[7.] If the sum of squared distances stabilizes over steps 2--6, then
the mean spike train is the current estimate and we can stop the
procedure. Otherwise, go back to
step~2.
\end{enumerate}

Denote the estimated mean in the $m$th iteration as $S^{(m)}$. One
can show that the sum of squared distances (SSD), $\sum_{k=1}^K
d_{\mathrm{GVP}}[\lambda](S^k,
S^{(m)})^2$, decreases iteratively as a function of $m$. As $0$ is a
natural lower bound, the SSD will always converge when $m$ gets large.
The detailed proof is given in the \hyperref[appen]{Appendix}. Note that this MAPC
algorithm takes only linear computational order on the number of spike
trains in the set. In practical applications, we find that this
algorithm has great efficiency in reaching a reasonable convergence to
a mean spike train.

\begin{figure}[b]

\includegraphics{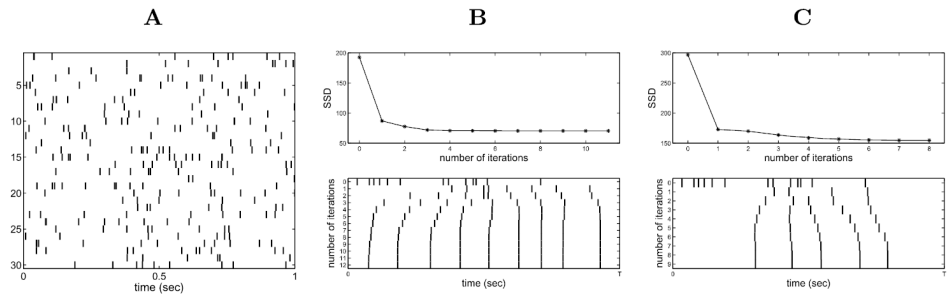}

\caption{\textup{A}: 30 spike trains generated from
a homogeneous Poisson process. Each vertical line indicates a
spike. \textup{B}:
Estimation results when $\lambda^2 = 6$. Upper panel:
The sum of squared distances (SSD) over all iterations.
Lower panel: The estimated mean spike train over all
iterations. The initial is the spike train on the top row (0), and
the final estimate is the spike train on the bottom row (12th).
\textup{C}:~Estimation result when $\lambda^2 = 60$.}
\label{fig:hpp}
\end{figure}

In general, when the penalty coefficient $\lambda$ gets large, the
optimal time warping chooses to have fewer matchings between spikes
to lower the warping cost. Some of the spikes in the mean will be
removed during the iterations to
minimize the SSD. In the extreme case, when $\lambda$ is sufficiently
large, any time warping would be discouraged (as that will result in a
larger distance than simply the Exclusive OR operation). In this case,
the mean spike train will be an empty train.
This result indicates that in order to get a meaningful estimate of the
mean spike train,
the penalty coefficient $\lambda$ should not take a very large value.
In practical use, one may use a cross-validation to decide the optimal
value of $\lambda$.

\subsubsection{Illustration of the MAPC algorithm}\label{sec2.4.2}

To test the performance of the MAPC algorithm, we illustrate the mean estimation
using 30 spike trains randomly generated from a homogeneous Poisson point
process. Let the total time $T = 1$~(sec), the Poisson rate $\rho= 8$~(spikes$/$sec). The individual spike trains are shown in Figure~\ref{fig:hpp}A. The number of spikes in these trains
varies from 3 to 13, with the median value of 9.
Therefore, $n$, the
number of spikes in the mean, is initialized to be 13 and we adopt
randomly distributed 13 spikes in $[0, T]$ as the initial for the mean
in each case.
We let $\lambda^2$ vary between 6 and 60 to show the behavior for a
small and a large warping penalty.

The result for the MAPC algorithm for $\lambda^2 = 6$ is shown in Figure~\ref{fig:hpp}B. The upper panel shows the evolution of the SSD in
equation (\ref{eq:mean})
versus the iteration index. We see that it takes only a few iterations for
the SSD to decreasingly converge to a minimum. The estimated mean spike
trains over iteration are shown in the lower panel. Apparent changes
are observed in the first few (1 to 5) iterations, and then the process
stabilizes. Note that the spikes in the mean train are approximately
evenly spaced, which properly
captures the homogeneous nature of the underlying process.
We also note that the number of spikes in this mean spike train is 9,
which equals the median of the numbers of spikes in the set.

The result for $\lambda^2 = 60$ is shown in Figure~\ref{fig:hpp}C.
With a larger penalty, the
optimal time warping between spike trains chooses to have fewer
matchings between spikes
to lower the warping cost. Some of the spikes in the mean are removed
during the iteration. In this case, the convergent SSD is about 150,
which is greater than the SSD when $\lambda^2 = 6$ (at about 80).
Note that when $\lambda$ is even larger, we expect fewer or even no
spikes to appear in the estimated mean.

\subsection{Advantages over previous methods of averaging spike trains}
\label{subsec:comparison}

%
\begin{table}
\tabcolsep=0pt
\caption{Comparison of average of spike trains for different methods}\label{tab:comp}
\begin{tabular*}{\tablewidth}{@{\extracolsep{\fill}}@{}lccccc@{}}
\hline
                 &                     & \textbf{``consensus''}        & \textbf{``prototype''}               & \textbf{``average''}        & \textbf{``mean''} \\
                 & \textbf{``mean''}   & \textbf{Victor and}         & \textbf{Diez, Schoen-}               & \textbf{Julienne and}       & \textbf{Wu and} \\
                 & \textbf{(proposed}  & \textbf{Purpura}            & \textbf{berg and}                    & \textbf{Houghton}           & \textbf{Srivastava} \\
\textbf{Method } & \textbf{framework)} & \textbf{(\citeyear{vpk97})} & \textbf{Woody (\citeyear{Diez2012})} & \textbf{(\citeyear{JMN13})} & \textbf{(\citeyear{WuSrivastavaJCNS11})} \\
\hline
Metric used & GVP metric &Victor--  & Victor-- & van Rossum  & Elastic  \\
&& Purpura metric & Purpura metric & metric & metric\\[3pt]
Properties (in  & 1, 2, 3, 4, 5 &2, 4, 5 & 2, 4, 5 & 1, 2, 4 & 1, 2, 5\\
\quad \hyperref[sec1]{Introduction})\\
\quad satisfied  \\[3pt]
Domain &Full spike  & Given   & Given  &Full spike  & Full spike  \\
& train space & sample & spike times & train space & train space\\
&& set & set\\[3pt]
Number of  spikes, & Median of  &NA&NA&NA & median of  \\
\quad $\lambda^2 \ll1$ & $\{n_1, \ldots, n_N\} $ & && &$\{n_1, \ldots, n_N\}$ \\[3pt]
Spike times if  & $c_j=\frac {1}{N} \sum_{k=1}^N s_{kj}$ & Restricted  & Restricted  & NA & ISI-based \\
\quad $n_1 = \cdots= n_N$, && to sample & to sampled && nonlinear \\
\quad $\lambda^2 \ll1$ && set & spike times & & form\\[3pt]
Uniqueness in the &Almost surely & Nonunique & Nonunique& Not known & Almost  \\
\quad full space && &&&surely\\
\hline
\end{tabular*}
\end{table}

There have been multiple ideas of capturing the general trend in a set
of spike trains, which include the ``consensus'' spike train [\citet
{vpk97}], the ``prototype'' spike train [\citet{Diez2012}],
the ``average'' spike train [\citet{JMN13}], and the ``mean'' spike
train [\citet{WuSrivastavaJCNS11}]. However, none of those concepts
satisfies all desirable 5 properties of a mean spike train listed in
the \hyperref[sec1]{Introduction} section.
We have summarized the most relevant differentiating features in Table~\ref{tab:comp}. In the case of the ``consensus'' and ``prototype''
spike trains, one main problem lies in the nonuniqueness of the results
in the spike train space, which arises directly from the underlying
metric used (resembles Manhattan distance). If the estimated spiking
times of those averages are restricted to spiking times in the sample
sets, then the estimates can be unreliable, particularly when sample
sizes are relatively small. The ``average'' design uses the van Rossum
metric, which relies on kernel-smoothing of the spike trains. The
estimation of the ``average'' is based on a greedy algorithm, but the
accuracy and the kernel dependence of the method have not been carefully examined. In
this article, we propose a new notion of a ``mean'' spike train based
on the kernel-free GVP metric. The key advantages behind our design are
the Euclidean properties of GVP distance and the subsequent Karcher
mean definition [in equation (\ref{eq:mean})]. The new framework
satisfies all 5 desirable properties, which distinguishes it from others.

It is worth noting that, to the best of our knowledge, the GVP metric
is one of only two spike train metrics that have Euclidean properties
[the other one is the Elastic metric proposed in \citet
{WuSrivastavaJCNS11}]. However, the Elastic metric satisfies only 3 out
of the 5 properties, and the GVP metric has two apparent advantages
over it: first, the Elastic metric estimates the mean spike train
explicitly depending on recording intervals. Such dependence may
introduce an additional level of noise arising from experimental
parameters, making the inference less reliable. In contrast, the mean
spike train under the GVP metric relies only on exact spike times in
the given data and is independent of recording intervals (Property~4).
Second, the fact that the Elastic mean is estimated through the
inter-spike intervals (ISI) makes it difficult to capture the intuition
behind the result, whereas the GVP mean is estimated directly through
spike times and resembles the intuition of the mean estimation
(Property~3).

\begin{figure}

\includegraphics{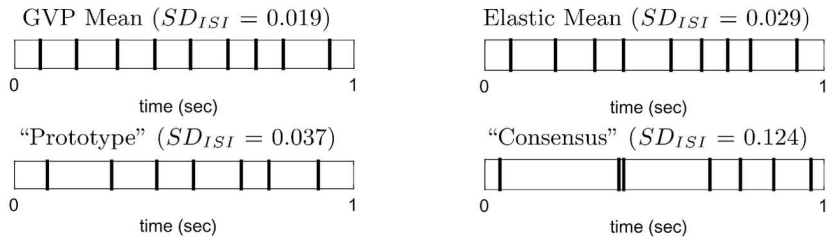}

\caption{Averaged spike trains according to four different methods.}
\label{fig:comp}
\end{figure}

For illustrative purposes, we compare spike train ``averages'' of all
methods using the 30 spike trains in Section~\ref{sec2.4.2}, where the data is
simulated under a homogeneous Poisson process. A natural expectation is
that these averages should be \textit{equi-distantly} spaced across the
time domain. We adopt a simple measure for the equi-distant spacing---compute the standard deviation of the ISI in each train, denoted by
$SD_{\mathrm{ISI}}$. Basically, smaller $SD_{\mathrm{ISI}}$ values indicate more even
spacing. In Figure~\ref{fig:comp}, we show the averages estimated
using the GVP mean, Elastic mean, ``Prototype,'' and ``Consensus''
methods. (In the GVP and Elastic methods, we let penalty coefficients
be sufficiently small.) It is found that $SD_{\mathrm{ISI}}$ in the GVP mean is
0.019, the smallest over all four methods.

%

\section{Application in noise removal}
\label{sec:application}

The notion of the mean spike train has a direct application in neural
decoding. In this article we examine how the mean can be used to remove
spontaneous activity in geniculate ganglion neurons and improve
decoding performance.

\subsection{Noise-removal method}
\label{subsec:nrm}
Geniculate ganglion neurons exhibit spiking response to the chemical
stimulus applied on the taste buds on the tongue. Such neuronal
activity is commonly used in the neural coding in the peripheral
gustatory system [\citet{DiLorenzo09,Breza10,LawhernFIN11}]. We note
that these neurons exhibit responses even if there is no stimulus
applied or the stimulation is a control solution---artificial saliva.
That is, the observed spike trains under the stimulation are likely to
be a mixture of the spontaneous activity and responses to the taste
stimuli. In the context of neural decoding, such spontaneous activity
can be viewed as ``background noise,'' and a ``de-noised'' spiking
activity is expected to better characterize the neural response with
respect to the taste stimulus and result in better decoding performance.

%
\begin{figure}

\includegraphics{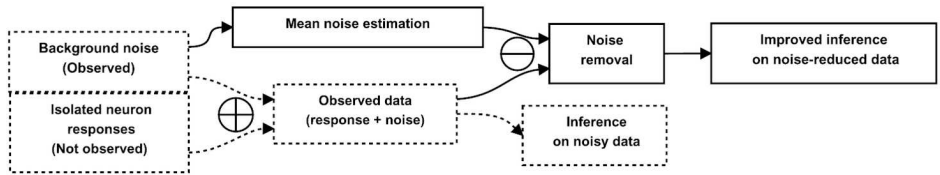}

\caption{Scheme differentiating the noise-removal approach from
standard inference on spike train data. Dashed boxes indicate the
components of standard inference framework, the solid lines indicate
where the noise-removal framework is introduced.}
\label{fig:flow}
\end{figure}

Previous approaches to the noise-removal focus mainly on spike count
across a time domain and do not have a temporal matching between
spikes. In this paper, we propose a novel noise-removal procedure based
on our new framework. In Figure~\ref{fig:flow} we describe the
schematic idea of incorporating the noise removal in statistical inference.
The procedure assumes that the observed data is a ``sum'' of isolated
neuron responses and their spontaneous activity. To improve the neural
decoding, we at first use the stimulus-free spike recordings to
estimate the mean background noise with the MAPC algorithm. Then, we
``subtract'' the mean out from the observed stimulus-dependent data.
Obviously, for random variables $X, Y$ in vector spaces one cannot
assume that $\tilde X= X + Y - {\bar Y}$ ($\bar Y$ denotes the mean of
$Y$) is a noise reduced ``version'' of $X+Y$. However, in the space of
spike trains we managed to establish this procedure, utilizing the
warping matchings on the GVP mean. The procedure of obtaining $\tilde
X= X \oplus Y \ominus\bar Y$ indeed gives a noise-reduced version
$\tilde X$ of a point pattern $X \oplus Y$. This approach is possible
with definitions of the addition $\oplus$ and the subtraction $\ominus
$ in the spike train space as follows:

\subsubsection*{Adding spike trains} We assume that the noise is additive and
adding spike trains is achieved by union set operation. That is, let $X
= (x_1, \ldots, x_N)$ and $Y =(y_1, \ldots, y_M)$ be two spike trains
of length $N$ and $M$, respectively. We define a spike train $Z = X
\oplus Y$ as a spike train of length $N+M$ such that
\[
Z=X \oplus Y = \bigl(\{ x_1, \ldots, x_N\} \cup
\{y_1,\ldots, y_M \} \bigr).
\]

\subsubsection*{Subtracting spike trains} Defining the subtraction is more
challenging, as it cannot follow directly from the set operations. This
is due to the fact that it is unlikely to have coinciding spike times
in two different spike trains. To perform the subtraction, we turn to
the definition of the GVP metric and optimal warping between two spike
trains [equation (\ref{eq:gvp})]. We define the subtraction of a spike
train $Y$ from a spike train $X$ as removing all spike positions from $X$
that are matched with spikes in $Y$ under the optimal warping $\gamma$.
We say that a spike pair $(x_i, y_j)$ is matched if $x_i = \gamma
(y_j)$ for some pair $(i,j)$.

Formally, let $X = (x_1, \ldots, x_N), Y =(y_1, \ldots, y_M) $ be two
spike trains and $\gamma$ be the optimal warping between them
according to the $d_{\mathrm{GVP}}$ metric. We define the subtraction of $Y$
from $X$, noted by $Z = X \ominus Y$, as follows:
\[
Z = X\ominus Y = \bigl(\{x_1, \ldots, x_N\}\setminus
\bigl\{ x_i : x_i = \gamma (y_j)\mbox{ for
some pair } (i,j) \bigr\} \bigr).
\]

Once the $\oplus$ and $\ominus$ are established, we can describe the
noise-removal method as follows: we ``subtract'' the $\operatorname{mean}(Y)$ from the
observed $X\oplus Y$ using the matching of the GVP metric and obtain a
spike train $X\oplus Y \ominus \operatorname{mean}(Y)$ by removing matched spikes.
We at first use a simulation for illustration of both $\oplus$ and
$\ominus$ operations in Section~\ref{sec:simulated}. The new method
is then applied to a real experimental data set in Section~\ref{sec:real}.

\subsection{Result for simulated data}
\label{sec:simulated}

%
\begin{figure}[b]

\includegraphics{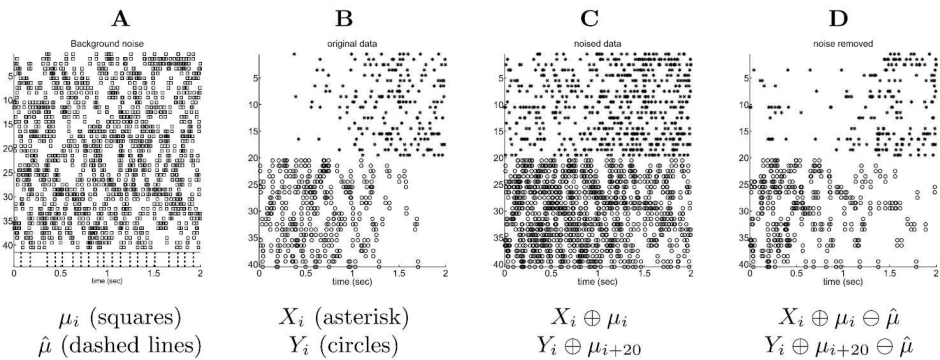}

\caption{Illustration of the noise addition and the noise removal with
the use of the $\oplus, \ominus$ operations. \textup{A}: Background
noise---40 spike trains generated from $\operatorname{HPP}(10)$; the mean background
noise is presented with dashed lines in the bottom row. \textup{B}:
$2\times 20$ spike trains from $\operatorname{IPP}(\rho_X)$ (asterisks) and $\operatorname{IPP}(\rho
_Y)$ (circles), respectively. \textup{C}: Sum of spike trains from \textup{A}
and \textup{B}. \textup{D}: Spike trains after the background noise removed.}
\label{fig:3figs_data}
\end{figure}

To illustrate the noise-removal framework, we first generate 40
independent realizations $\{\mu_i\}_{i=1}^{40} $ of a homogeneous
Poisson process on $[0, 2]$ with constant intensity $\alpha=10$. These
simulations represent the noise and are used to estimate the mean
background noise $\hat\mu$ with the MAPC algorithm. The results are
shown in Figure~\ref{fig:3figs_data}A.

Next we generate two sets of $20$ independent spike trains, $\{X_i\}
_{i=1}^{20}$ and $\{Y_i\}_{i=1}^{20}$, as realizations of an
inhomogeneous Poisson processes (IPP) with intensity functions $\rho
_X(t) = \exp (-(t-1.5)^2)$ and $\rho_Y(t) = \exp (-(t-0.5)^2)$,
respectively. The generated spike trains are shown in Figure~\ref{fig:3figs_data}B. In our framework they correspond to the underlying
true neuronal signals.

In the third step we obtain the equivalent of the ``observed'' data, by
adding the previously generated noise for each generated $\mu_i$ to
the corresponding spike trains $X_i$ and $Y_i$. The combined results
are shown in Figure~\ref{fig:3figs_data}C. In this case adding spike
trains is understood in the set operations terms. We obtain spike
trains following Poisson processes $X_i \oplus\mu_i \sim \operatorname{IPP}(\rho_X
+ \alpha), Y_i\oplus\mu_{i+20} \sim \operatorname{IPP}(\rho_Y + \alpha)$.

The mean background noise spike train $\hat\mu$ is then subtracted
out from each realization of the noised data set, according to the
procedure described in Section~\ref{subsec:nrm}. For each $i = 1,
\ldots,20$, we obtain the noise removed spike trains: $X_i \oplus\mu
_i \ominus\hat\mu, Y_i\oplus\mu_{i+20} \ominus\hat\mu$,
shown in Figure~\ref{fig:3figs_data}D.

%
\begin{figure}

\includegraphics{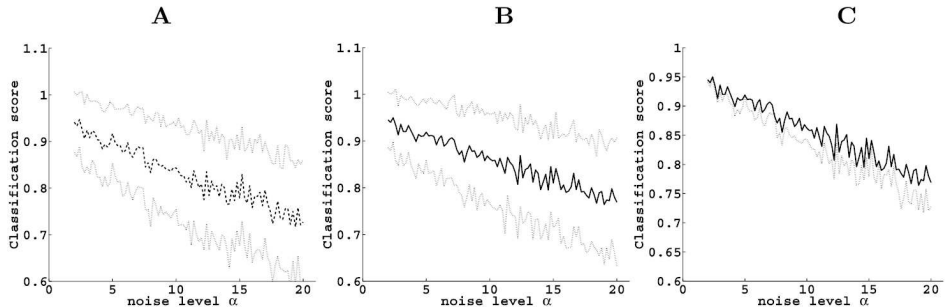}

\caption{The noise-removal influence on classification performance
with respect to increasing noise level $\alpha$. \textup{A:} The
classification performance for the noisy data. The bold lines represent
the average classification score among 50 simulations, the dotted lines
indicate the standard deviation from the average classification score.
\textup{B} Same as \textup{A}, but for the noise-removed data. \textup{C}: Mean
classification score curves from \textup{A} (dashed line) and \textup{B}
(dotted line).}
\label{fig:3figs_score}
\end{figure}

We repeat this simulation procedure 50 times for each level of $\alpha
\in[2,20]$ and perform classification on the noisy data $X_i \oplus
\mu_i, Y_i\oplus\mu_{i+20}$ as well as on the noise-removed: $X_i
\oplus\mu_i \ominus\hat\mu, Y_i\oplus\mu_{i+20} \ominus\hat
\mu$. The classification score is obtained by a standard leave-one-out
cross-validation. We record the average score (the classification
accuracy) with the standard deviation for each $\alpha$ level; the
result is shown in Figure~\ref{fig:3figs_score}A,~B. As anticipated with
the increasing noise intensity $\alpha$, the classification
performance on each of the noisy and noise-reduced data sets declines.
However, if we compare the two average classification scores presented
in Figure~\ref{fig:3figs_score}C, we see that the noise-removal
framework outperforms the classification on the noisy data once the
noise intensity level $\alpha$ becomes not negligible. This result
indicates that the proposed noise-removal procedure can help increase
the contrast between different classes and result in an improvement in
classification analysis. Next, we will examine this procedure on a real
experimental data set.

\subsection{Result in real data in gustatory system}
\label{sec:real}

%
\begin{figure}

\includegraphics{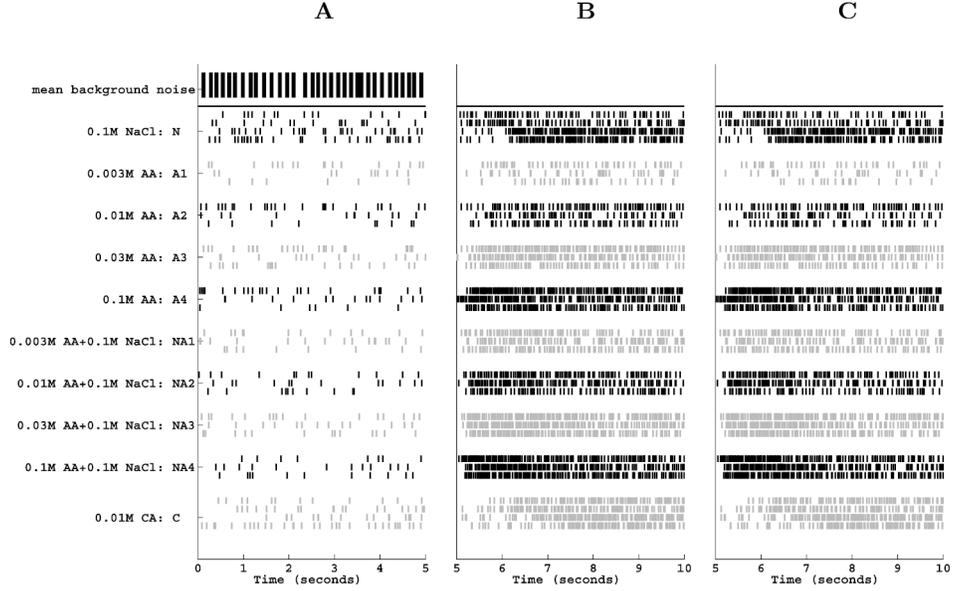}

\caption{An example of spike trains from Cell 10. Each group of 3 or 4
rows corresponds to a different type of stimuli applied. \textup{A}: The
5-second pre-stimulus spike trains, whose mean spike train, calculated
by the MAPC algorithm, is shown by the thick vertical bars on the top
of the panel. \textup{B}: The 5-second stimulus period. \textup{C}: The same
5-second period of spike trains as in \textup{B}, but with spontaneous
activity subtracted out.}\label{fig:realData}
\end{figure}

Here we apply the noise-removal procedure to neural response in the
gustatory system and test if the decoding
(i.e., classification with respect to taste stimuli) can improve after
the spontaneous activity is removed.
The data consists of spike train recordings of rat geniculate ganglion
neurons and was previously used in \citet{WuPLOSONE13}.
Briefly, adult male Sprague--Dawley rat's geniculate ganglion tongue
neurons were stimulated with 10 different solutions over a time period
of 5 seconds: 0.1 M NaCl, 0.01 M citric acid (CA), 0.003,
0.01, 0.03, and 0.1 M acetic acid (AA), and each AA mixed
with 0.1 M NaCl. Each stimulus was presented 2--4 times.
Stimulus trials were divided into three time regions: a 5-second pre-stimulus
period, a 5-second stimulus application period, and a 5-second post-stimulus
period. During the first and third regions, a control solution of
artificial saliva was applied. During the stimulus period, one of the
10 aforementioned solutions was applied. In this study we focus on
classifying the given spike trains according to the 10 stimuli
presented in each of 21 observed neurons. In Figure~\ref{fig:realData}A,~B we present the real data recordings in the first and
second time regions from one example neuron.

The spike trains in the pre-stimulus 5-second period reflect
spontaneous activity with artificial saliva applied. They are treated
as ``noise'' data, in contrast to stimulus-dependent responses. We
compute their mean spike train with the parameter $\lambda^2 = 0.001$
(a small value to get more spikes in the mean). The result is shown in
the top row in Figure~\ref{fig:realData}A.
This mean properly summarizes spiking activity during the pre-stimulus period.
In the next step, we subtract out this mean noise from the data during
the stimulus period (spike trains between the 5th and 10th second). The
noise-removed spike trains are shown in Figure~\ref{fig:realData}C.

We can now compare the decoding performance using the observed\break 
stimulus-response data and the ``noise-removed'' data. To reliably
evaluate classification scores, we take the approach of leave-one-out
cross-validation. In both cases of the observed data and the
noise-removed data, the class is assigned according to the nearest
neighbor's class under the $d_{\mathrm{GVP}}$ metric. In this classification
analysis, we use $\lambda^2=225$, a relatively larger value to
emphasize the importance of both matching term and penalty term in
equation (\ref{eq:gvp}).

%
\begin{figure}

\includegraphics{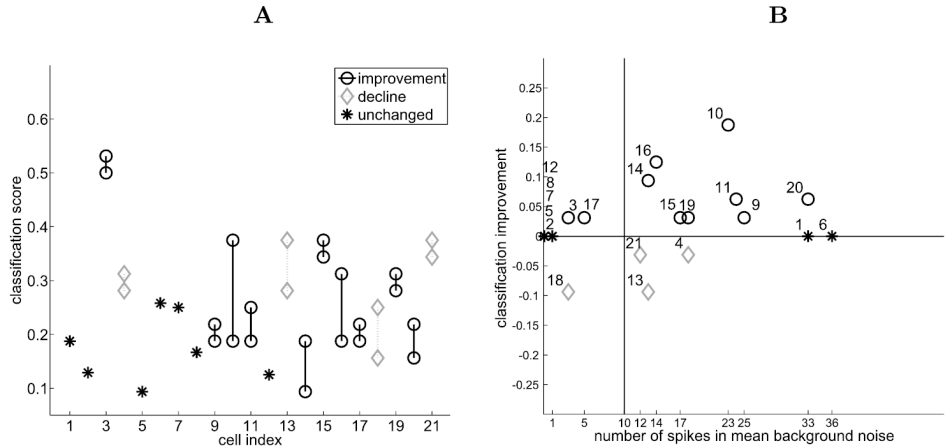}

\caption{The result of the noise-removal procedure applied to each of
the recorded 21 cells. The marker coding is the same for both panels
and indicates the influence of the noise-removal approach on the
classification score: black circles---increase, grey diamonds---decrease, black
asterisks---unchanged. \textup{A.} Raw classification
scores for each cell in each condition. \textup{B.} Same result as in \textup{A},
but in terms of classification score increase with respect to mean
noise size. The vertical black line corresponds to the noise size
cutoff of 10 spikes.}
\label{fig:realResults}
\end{figure}

The comparison on the classification accuracy is shown in Figure~\ref{fig:realResults}A. In 10 out of 21 cells the classification was
improved after the noise-removal procedure and in only 4 cells the
classification was hindered. Classification in 7 cells remained
unchanged after noise removal, which seems to be quite significant. To
explain this issue, we have investigated the size of the mean
background noise and its influence on increase in classification
performance. It turned out, as seen in Figure~\ref{fig:realResults}B,
that in 5 out of these 7 unchanged cells, the pre-stimulus spiking is
negligible (the estimated mean spike train has 0 or 1 spike). In those
cases, obviously, subtracting out the mean noise spike train will bare
minimum influence.

The remaining two cases are associated with the opposite problem of the
noise size---the number of spikes in the pre-stimulus period is
comparable or greater than the number of spikes in the stimulation
period. When such mean noise is subtracted out, it also can take away
relevant information, thus it may not improve the decoding. Those noise
size issues are consistent with common intuition behind the noise
removal. However, it is worth noting that the size of the estimated
mean background noise can be controlled in our new framework, by
adjusting the penalty parameter $\lambda$ (Section~\ref{sec:comp_mean}). More investigation will be conducted on the selection
of $\lambda$ in our future work.

When focusing on cells that have significant noise influence in their
spiking pattern (at least 10 spikes in the estimated mean noise spike
train), we see that in the majority of cases (8 out of 13) the noise
removal has improved the classification score (Figure~\ref{fig:realResults}B).
In extreme cases we obtain up to 20$\%$ improvement in the decoding
performance. (Note that with 10 different stimuli, a random guess
results in 10\% average accuracy.) Only 3 out of the 13 cells indicate
loss of information and 2 are not influenced by noise removal.

In summary, we find that our notion of mean background noise for spike
train data is in agreement with the common understanding of the
additive noise for the majority of recorded neurons. Moreover, the
proposed noise-removal framework effectively improves neural decoding,
provided that the pre-stimulus spiking has a high enough intensity.

\section{Discussion}
\label{sec:discussion}

In this article we propose a new framework for defining the mean of a
set of spike trains and the deviation from the mean. We provide an
efficient algorithm for computing the mean spike train and prove the
convergence of the method. The framework is based on the $d_{\mathrm{GVP}}$
metric [\citet{Dubbs10}] which resembles the Euclidean distance. This
concept gives an intuitive sample mean point pattern, in which the
spike positions in the mean are averaged among matched spikes in a set
of all spike trains.

Our summary statistics provide the basis for inference on point pattern
data and in this article we utilize it to develop a mean-based noise-removal approach. We show that our procedure improves the
classification score for simulated inhomogeneous Poisson point process
data with various nonnegligible noise levels. We have also applied the
new tools to a neural decoding problem in a rat's gustatory system. It
is found that the mean point pattern approach and the noise-removal
framework significantly improved the neural decoding among the set of
21 neurons.

In the noise-removal framework, we have defined the operations of
addition and subtraction between spike trains with the use of the
matching component of the GVP metric. We, however, note that those
operations do not satisfy the law of associativity. For more advanced
analysis, it is desirable to establish an algebraic structure on the
space of point patterns. Thus, refining those approaches will be
pursued in the further work.

Once the algebraic structure is established, statistical models can be
built and regression analysis can be performed.
With this setting and the already developed mean and the deviation from
the mean approaches, we expect to develop classical statistical
inferences such as hypothesis tests,
confidence intervals/regions, FANOVA (functional ANOVA), FPCA (functional
PCA), and regressions on functional data
[\citet{ramsay-silverman-2005,Valderrama07}]. All these tools are
expected to provide a new methodology for more effective analysis and
modeling of neural spike trains or any point pattern data in general.


\begin{appendix}
\section*{Appendix}\label{appen}

%
\begin{theorem}[(Convergence of the MAPC algorithm)]
Denote the estimated mean in the $m$th iteration of the MAPC-algorithm
as $S^{(m)}$. Then the sum of squared distances
$\sum_{k=1}^K d_{\mathrm{GVP}}(S^k, S^{(m)})^2$ decreases iteratively. That is,
\[
\sum_{k=1}^K d_{\mathrm{GVP}}
\bigl(S^k, S^{(m+1)} \bigr)^2 \le\sum
_{k=1}^K d_{\mathrm{GVP}} \bigl(S^k,
S^{(m)} \bigr)^2.
\]
\end{theorem}

\begin{pf}
The\vspace*{1pt} proof will go through steps 2--6 of the algorithm---in
each step we will show that the overall distance to the proposed mean
$S^{(m)} = (s_1^{(m)}, \ldots, s_n^{(m)})$ is nonincreasing:
\begin{longlist}
\item[1. \textit{Matching}:] In the $m$th iteration, we find the optimal
matching $\gamma^k$ from $S^{(m)}$ to $S^k$ for each $k \in\{1,\ldots,
K\}$. Having those, we can write
\begin{eqnarray*}
&& \sum_{k=1}^K d_{\mathrm{GVP}}
\bigl(S^k, S^{(m)} \bigr)^2
\\
&&\qquad = \sum
_{k=1}^K E_{\mathrm{OR}} \bigl(S^k,
\gamma^k \bigl(S^{(m)} \bigr) \bigr) + \lambda^2
\sum_{k=1}^K\sum
_{\{i,j: s_i^k =
\gamma^k(s^{(m)}_j)\}} \bigl(s^k_i -
s^{(m)}_j \bigr)^2.
\end{eqnarray*}

\item[2. \textit{Adjusting}:] By definition, we update $S^{(m)}$ to $\tilde
S^{(m)} = (\tilde s_1^{(m)}, \ldots, \tilde s_n^{(m)}) = \frac
{1}{K}\sum_{k=1}^K R_k$, where
$R_k = (r_1^k, \ldots, r_n^k)$ with
%
\begin{eqnarray}
r_j^k = \cases{ s_i^k, &\quad
if $\exists i \in\{1, \ldots, n_k\}$, s.t. $\gamma ^k
\bigl(s_j^{(m)} \bigr) = s_i^k$,
\cr
s_j^{(m)}, &\quad otherwise}\nonumber
\\
\eqntext{k = 1, \ldots, K, j = 1, \ldots, n.}
\end{eqnarray}
Hence,
$\sum_{k=1}^K\sum_{\{i,j: s_i^k = \gamma^k(s^{(m)}_j)\}}(s^k_i - s^{(m)}_j)^2
= \sum_{k=1}^K\sum_{j=1}^n (r^k_j - s^{(m)}_j)^2
\ge\sum_{k=1}^N\*\sum_{j=1}^n (r^k_j - \tilde s^{(m)}_j)^2$.

Let $\gamma$ be the piecewise linear warping function from $S^{(m)}$
to $\tilde S^{(m)}$, that is, $\tilde S^{(m)} = \gamma(S^{(m)})$. Then
\begin{eqnarray*}
&& \sum_{k=1}^K d_{\mathrm{GVP}}\bigl(S^k, S^{(m)}\bigr)^2
\\
&&\qquad \ge \sum_{k=1}^K E_{\mathrm{OR}}\bigl(S^k, \gamma^k\bigl(S^{(m)}\bigr)\bigr) + \lambda^2 \sum_{k=1}^K\sum_{j=1}^n
\bigl(r^k_j - \tilde s^{(m)}_j\bigr)^2
\\
&&\qquad \ge \sum_{k=1}^K E_{\mathrm{OR}}\bigl(S^k, \gamma^k \bigl( S^{(m)}\bigr)\bigr) +
\lambda^2 \sum_{k=1}^K\sum_{\{i,j: s_i^k = \gamma^k( s^{(m)}_j)\}} \bigl(s^k_i - \tilde
s^{(m)}_j\bigr)^2
\\
&&\qquad = \sum_{k=1}^K E_{\mathrm{OR}}\bigl(S^k, \gamma^k \circ\gamma^{-1}\bigl(\tilde
S^{(m)}\bigr)\bigr) + \lambda^2 \sum_{k=1}^K\sum_{\{i,j: s_i^k = \gamma^k
\circ\gamma^{-1} \bigl(\tilde s^{(m)}_j\bigr)\}} \bigl(s^k_i - \tilde s^{(m)}_j\bigr)^2
\\
&&\qquad \ge\sum_{k=1}^K d_{\mathrm{GVP}}\bigl(S^k, \tilde S^{(m)}\bigr)^2.
\end{eqnarray*}

\item[3. \textit{Prunning}:]
$\tilde S^{*(m)} = \{ s_j \in\tilde S^{(m)} | \sum_{k=1}^K 1_{s_j \in
{\gamma^k(S^k)}} \ge K/2\}$ is a subset of $\tilde S^{(m)}$ in which
all spikes appear in $\{\gamma^k(S^k)\}_{k=1}^K$ at least $K/2$ times.\vspace*{1pt}
Based on the result in the Adjusting step, we have
$
\sum_{k=1}^K d_{\mathrm{GVP}}(S^k, S^{(m)})^2
\ge\break \sum_{k=1}^K E_{\mathrm{OR}}(S^k, \gamma^k \circ\gamma^{-1}(\tilde
S^{(m)})) + \lambda^2 \sum_{k=1}^K\sum_{\{i,j: s_i^k = \gamma^k
\circ\gamma^{-1} (\tilde s^{(m)}_j)\}} (s^k_i - \tilde s^{(m)}_j)^2$.
Using the basic rule of the Exclusive OR, it is easy to find that
\[
\sum_{k=1}^K E_{\mathrm{OR}}
\bigl(S^k, \gamma^k \circ\gamma^{-1} \bigl(
\tilde S^{*(m)} \bigr) \bigr) \le\sum_{k=1}^K
E_{\mathrm{OR}} \bigl(S^k, \gamma^k \circ
\gamma^{-1} \bigl(\tilde S^{(m)} \bigr) \bigr).
\]
Let $\tilde S^{*(m)} = (s_1^{*(m)}, \ldots, s_{n^*}^{*(m)})$, where
$n^*$ denotes the number of spikes in $\tilde S^{*(m)}$. Then,
\begin{eqnarray*}
&& \sum_{k=1}^K d_{\mathrm{GVP}}\bigl(S^k, S^{(m)}\bigr)^2
\\
&&\qquad \ge\sum_{k=1}^K E_{\mathrm{OR}}\bigl(S^k, \gamma^k \circ\gamma^{-1}\bigl(\tilde
S^{*(m)}\bigr)\bigr) + \lambda^2 \sum_{k=1}^K\sum_{\{i,j: s_i^k = \gamma^k
\circ\gamma^{-1} (\tilde s^{*(m)}_j)\}} \bigl(s^k_i - \tilde s^{*(m)}_j\bigr)^2
\\
&&\qquad \ge\sum_{k=1}^K d_{\mathrm{GVP}}\bigl(S^k, \tilde S^{*(m)}\bigr)^2.
\end{eqnarray*}

\item[4. \textit{Checking}:]
Finally, we perform the checking step to avoid the possible local
minima in the pruning process. In the test if a spike can be removed
from $\tilde S^{*(m)}$, we let $\hat{S}^{*(m)}$ be $\tilde S^{*(m)}$
except one spike with a minimal number of appearances. Then update the
mean spike train as
\[
S^{**(m)} = \cases{ \displaystyle\hat S^{*(m)}, &\quad if $\displaystyle\sum
_{k=1}^K d_{\mathrm{GVP}} \bigl(S^k,
\hat S^{*(m)} \bigr)^2 < \sum_{k=1}^K
d_{\mathrm{GVP}} \bigl(S^k, \tilde S^{*(m)}
\bigr)^2$,
\cr
\tilde S^{*(m)}, &\quad otherwise.}
\]
It is easy to verify that $\sum_{k=1}^N d_{\mathrm{GVP}}(S^k, S^{(m)})^2 \ge
\sum_{k=1}^N d_{\mathrm{GVP}}(S^k, S^{**(m)})^2$. In the test if a spike can be
added to $\tilde S^{**(m)}$, we let $\hat S^{**(m)}$ be $S^{**(m)}$
with one spike inserted at random within $[0, T]$. Then update the mean
spike train as
\[
S^{***(m)} = \cases{ \hat S^{**(m)}, &\quad if $\displaystyle\sum
_{k=1}^K d_{\mathrm{GVP}} \bigl(S^k,
\hat S^{**(m)} \bigr)^2 < \sum_{k=1}^K
d_{\mathrm{GVP}} \bigl(S^k, S^{**(m)} \bigr)^2$,
\cr
\tilde S^{**(m)}, &\quad otherwise.}
\]
\end{longlist}

\noindent
It\vspace*{2pt} is easy to see that $\sum_{k=1}^K d_{\mathrm{GVP}}(S^k, S^{(m)})^2 \ge\sum_{k=1}^K d_{\mathrm{GVP}}(S^k, S^{***(m)})^2$.
Using step 6, the mean at the $(m+1)$th iteration is $S^{(m+1)} =
S^{***(m)}$. Hence,
\[
\sum_{k=1}^K d_{\mathrm{GVP}}
\bigl(S^k, S^{(m+1)} \bigr)^2 \le\sum
_{k=1}^K d_{\mathrm{GVP}} \bigl(S^k,
S^{(m)} \bigr)^2.
\]\upqed
\end{pf}
\end{appendix}



%

\printaddresses
\end{document}